\documentclass{ttp_dsl2006}

\usepackage{graphicx}
\usepackage[intlimits]{amsmath}
\usepackage{amssymb}
\usepackage{exscale}
\usepackage{hyperref}
\usepackage{times}
\usepackage{subfigure}

\renewcommand{\large}{\fontsize{14}{18pt}\selectfont}
\renewcommand{\small}{\fontsize{11}{13.6pt}\selectfont}

\newcommand{\titleformat}{\sffamily\bfseries \large}						
\newcommand{\keywordsformat}{\noindent \small \sffamily}				
\newcommand{\abstractformat}{\noindent \textbf}						
\newcommand{\contentformat}{\rmfamily \normalsize \vspace{18pt}}			
\newcommand{\email}{\sffamily \small \vspace{-8pt}}						
\renewcommand{\subsection}{\textbf}	

\hypersetup{
    colorlinks,%
    citecolor=black,%
    filecolor=black,%
    linkcolor=black,%
    urlcolor=black
}

\begin{document}

\title{\titleformat  Clustering in generalized 1D Ising models }
\author{ Petr D. Andriushchenko\inst{1}$^{,\rm{a{\rm{*}}}}$ \textbf{,} and Konstantin V. Nefedev \inst{1}$^{,\rm{b}}$}
\institute{\sffamily $^{\rm 1}$Department of Computer Systems, School of Natural Sciences, Far Eastern Federal University, 690091, Vladivostok, Russia}

\maketitle

\begin{center}
\email{ $^{\rm a}$andriushchenko.pd@dvfu.ru, $^{\rm b}$nefedev.kv@dvfu.ru}
\end{center}

\keywordsformat{{\textbf{Keywords:}} magnetization, one-dimensional Ising model, order parameter, frustrations.}

\contentformat

\abstractformat{Abstract.} The research of one-dimensional Ising model with the different number of neighbors and the different sign of the exchange integral were performed. Magnetization shows the transition from disorder to order (in finite systems) only at positive exchange integral $J$ but clustered order parameter describes the transition regardless of the sign of the exchange integral and the number of neighbors.

\section{Introduction}
Today may seem from the side, that the theory of ordering reached its peak and now is reaping the fruits of their approaches on new structures, elaborating the critical indices and the phase transition temperature. But recently, theorists and experimenters publish the evidence of the opposite opinion that current approaches were ill-suited to describe the order in frustrated systems or systems with quenched disorder \cite{morgan2011thermal, kapaklis2014thermal,daunheimer2011reducing,morgan2013real,ahlberg2011two}. But to understand the processes occurring in such complex systems, you need to perfectly represent what happens in the simpler models. That is why finding a common approach to ordering need to start with simple models. Root mean square magnetization $<M^2>$ or its module is usually used to describe transitions in magnets models. This parameter can be used to study phase transitions in ferromagnets. However, as soon as we change the sign of the exchange integral with "+" to "-", the order parameter becomes 0 for any $T$ and $N\rightarrow \infty$. Therefore, the problem of determining the universal order parameter, especially for systems with a negative or alternating exchange interaction is an actual today \cite{antal2004probability,gambardella2002ferromagnetism}.

\begin{figure}[b!]
 \begin{center}
 \includegraphics[width=1\linewidth]{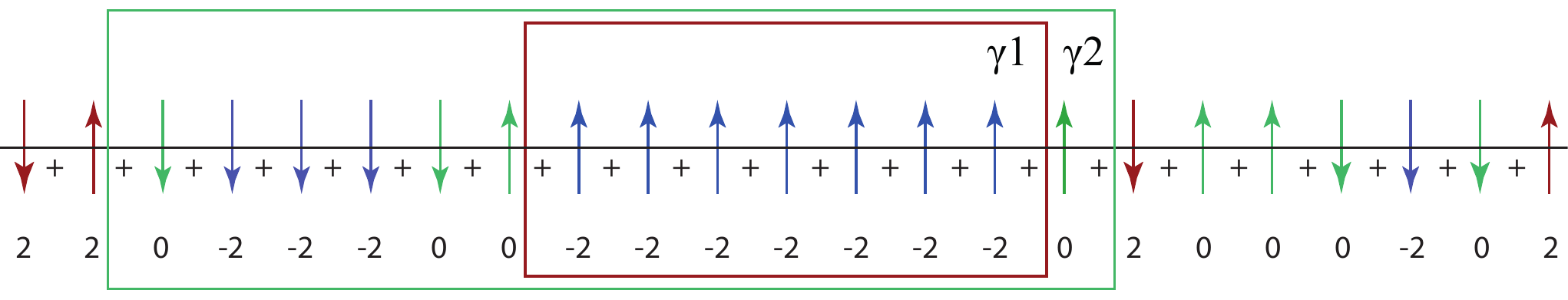}
 \end{center}
 \caption{An example of calculating the clustered order parameter on ferromagnet $(J > 0)$ 1D Ising spin chain with two nearest neighbors. Each spin in the chain has a different color (blue, green and red) depending on its interaction energy respectively $(-2,0,2)$. Red line highlights the biggest cluster of spins in the lowest energy state ($-2$, that is blue). The ratio of the spins in ground state in the maximal cluster to the total number of spins in a system is the order parameter $\gamma_1$. In this sample number of the spins in ground state in the maximal cluster is $7$, $n=23$, $\gamma_1=7/23\approx0.3$ }
   \label{Figure.cluster}
   \end{figure}

Therefore, we have proposed an approach to the determination of the order parameter to describe the transition from a disordered phase to an ordered (or partially ordered) \cite{andriushchenko2013magnetic}. Order parameter can be a ratio of the maximum size of the cluster formed by spins in the ground state (lowest energy) to the total number of particles (Fig.\ref{Figure.cluster}).
\begin{eqnarray}
\mathcal{H} = - \sum_{i=1}^{n}{\sum_{j=1}^z{J_{ij}S_iS_j}}-h\sum_{i=1}^n{S_i}
\label{eqq1}
\end{eqnarray}

And the simplest magnetic model is the one-dimensional Ising model \cite{fan2011one}. Classical 1D Ising model with Hamiltonian (\ref{eqq1}) was investigated, where $n$ - total number of spins, $J_{ij}$ - exchange integral and  $z$ - the number of nearest neighbors, which interact with spin. The model with $z = 2,3,4$ considered in this paper.

\section{1D Ising model with two nearest neighbors}


The 1D Ising model with $n=1000$ spins with two nearest neighbors and periodic boundary conditions (PBC) 
was investigated.

First of all, the ferromagnetic model with a positive exchange integral was calculated. From now on the simulation was performed with using Monte Carlo (MC), the number of MC steps for every $0.02T$ was $10^{10}$. As expected, the temperature dependence of the clustered order parameter (COP) $\gamma_1$ is almost identical to the behavior of magnetization. That is, in the case of ferromagnetic ($J > 0$) the COP $\gamma_1$ describes a transition from disordered phase to ordered like magnetization.

\begin{figure}[h]
\begin{minipage}[h]{0.49\linewidth}
\center{\includegraphics[width=1\linewidth] {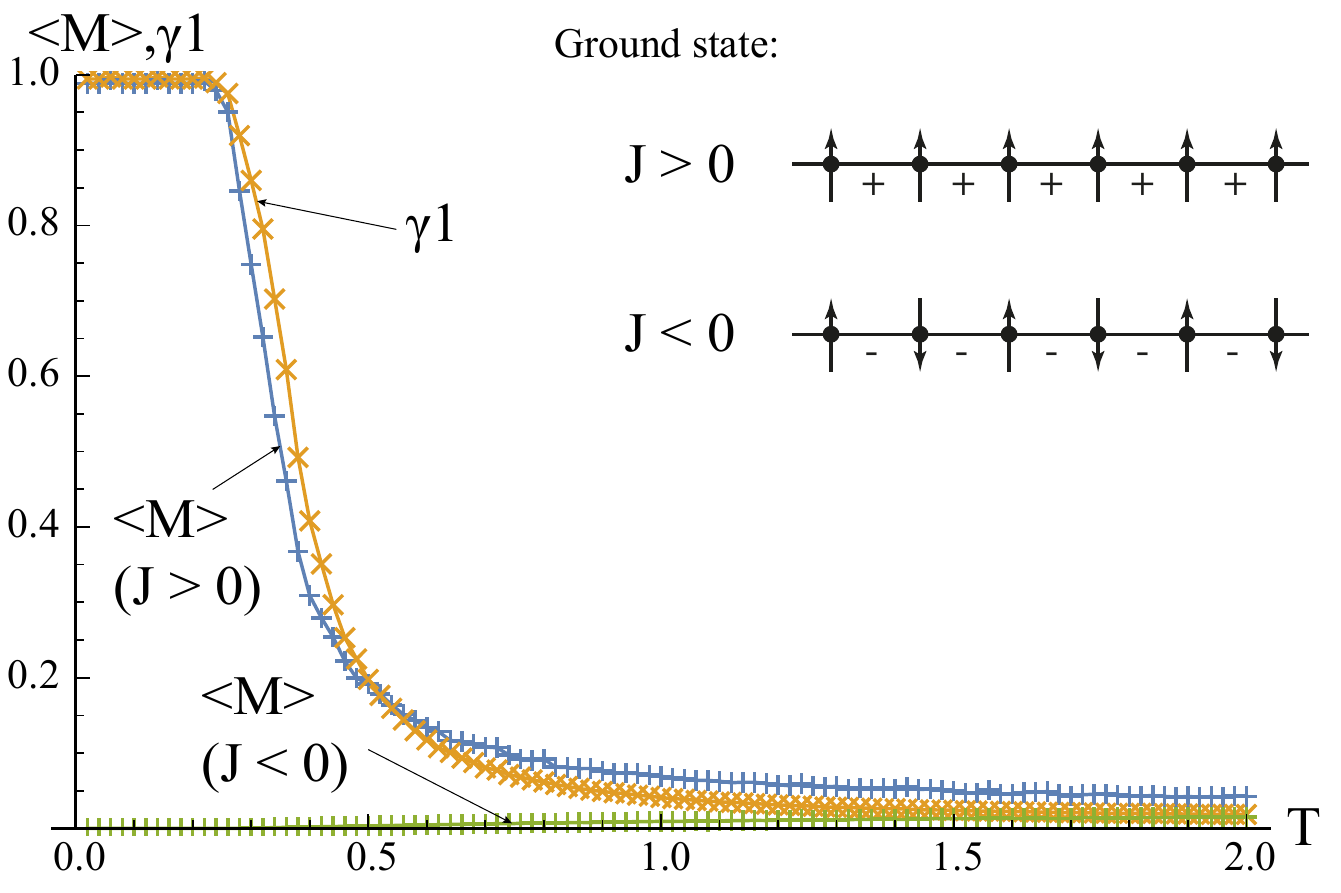}} \\ a)
\end{minipage}
\hfill
\begin{minipage}[h]{0.49\linewidth}
\center{\includegraphics[width=1\linewidth]{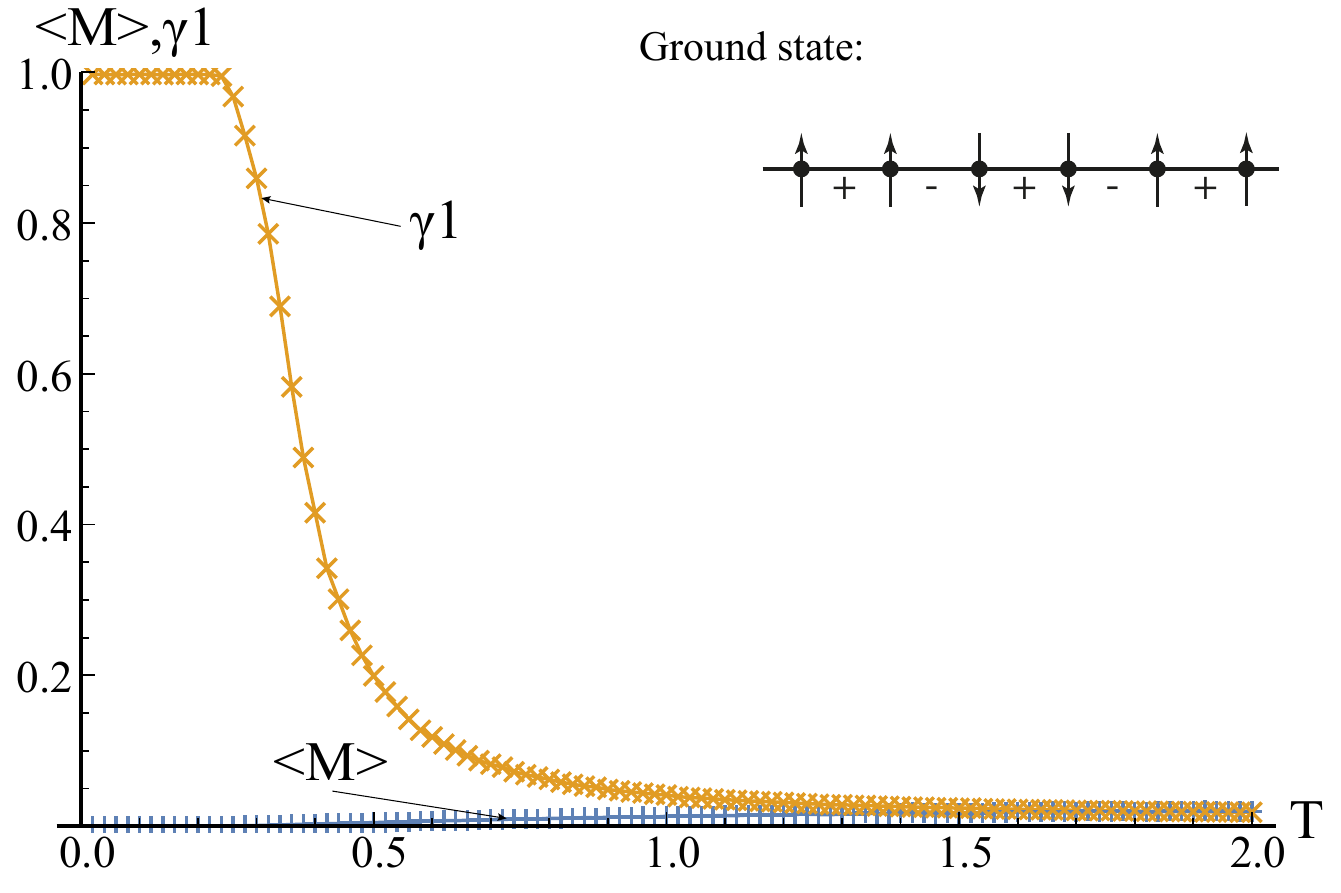}} \\ b)
\end{minipage}
\caption{The temperature behavior of COP $\gamma_1$ and magnetization of 1D Ising model (1000 spins) with two nearest neighbors and PBC. a) Blue denotes the mean magnetization of the ferromagnetic model with $J > 0$, green - magnetization of the antiferromagnetic model with $J < 0$, orange – COP (the same for these models). b) Blue denotes the magnetization with alternating exchange integral, orange - the same COP for such model.}
\label{Figure::1D2N_fig}
\end{figure}

If you change the sign of the exchange integral $J$ to the negative, the magnetization becomes zero at any temperature, as the ground state of the system will be antiferromagnetic (Fig. \ref{Figure::1D2N_fig}a). 

Despite this, COP $\gamma_1$ continues to describe the transition to the ordered phase (Fig.\ref{Figure::1D2N_fig}a).

If we consider a system with alternating exchange integral (each spin has one positive link and negative second one), it becomes clear that such system is not much different from the antiferromagnet, the magnetization of such system is also equal to 0 for any $T$, but the COP describes the ordering process (Fig.\ref{Figure::1D2N_fig}b).

\section{1D Ising model with three nearest neighbors}

\begin{figure}[h]
\begin{minipage}[h]{0.49\linewidth}
\center{\includegraphics[width=0.8\linewidth] {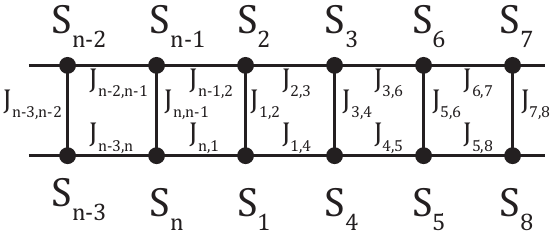}}
\end{minipage}
\hfill
\begin{minipage}[h]{0.49\linewidth}
\center{\includegraphics[width=0.6\linewidth]{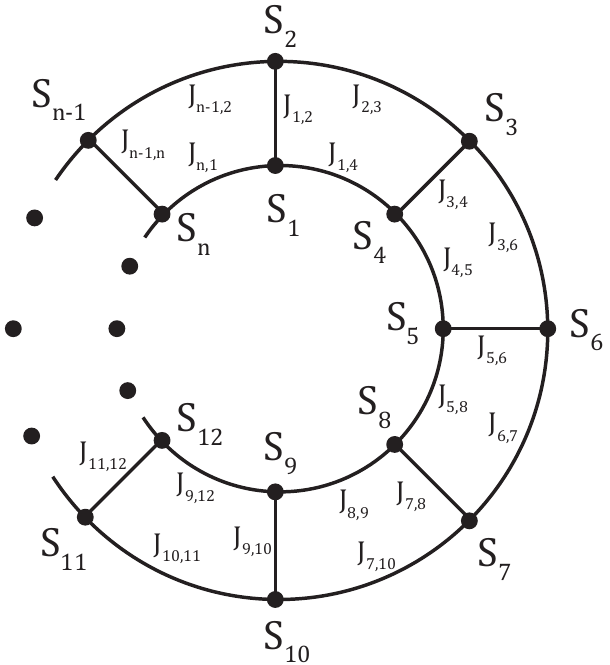}}
\end{minipage}
\caption{1D Ising model with three close neighbors and PBC.}
\label{Figure::1D3N}
\begin{minipage}[h]{0.49\linewidth}
\center{\includegraphics[width=1\linewidth] {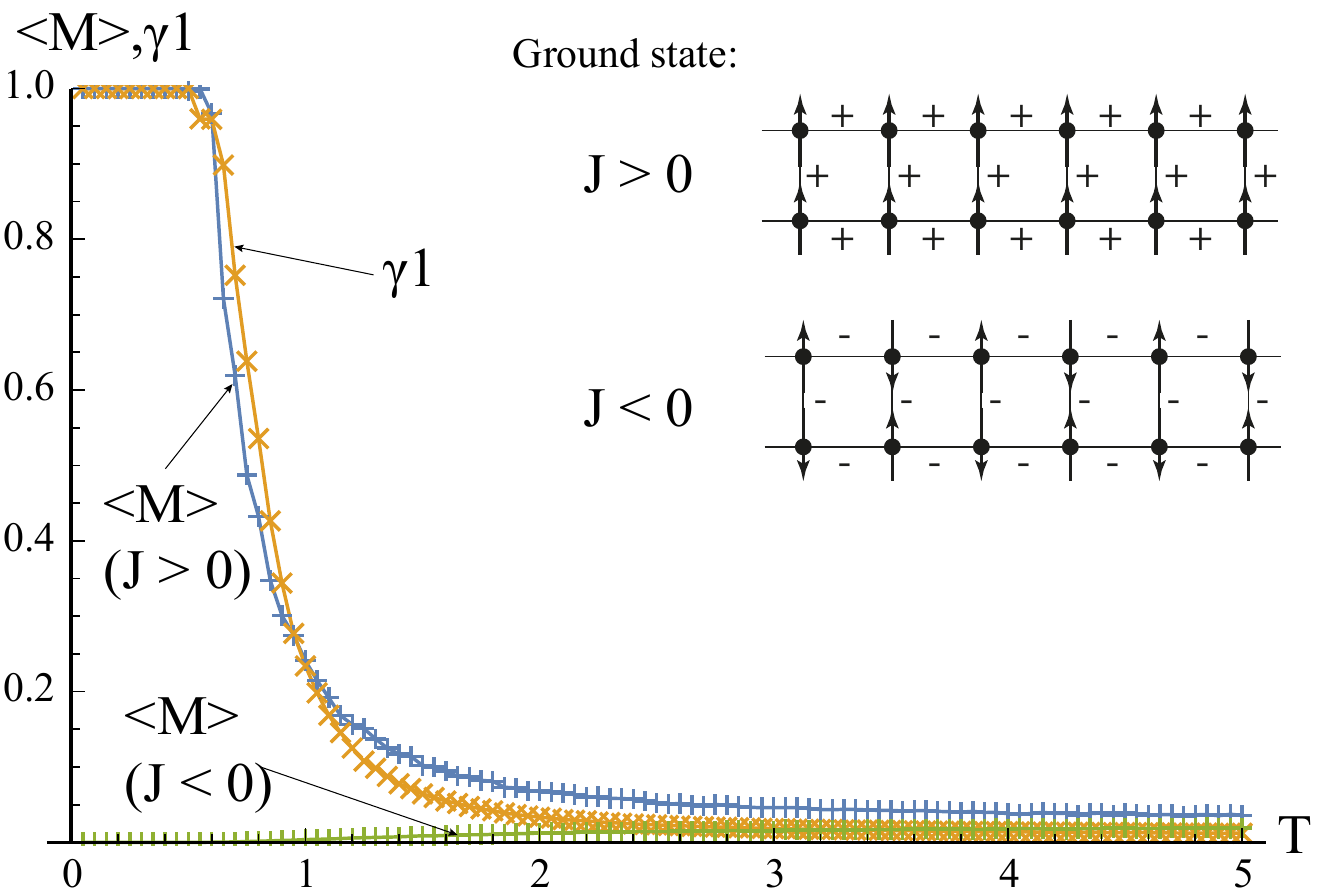}} \\ a)
\end{minipage}
\hfill
\begin{minipage}[h]{0.49\linewidth}
\center{\includegraphics[width=1\linewidth]{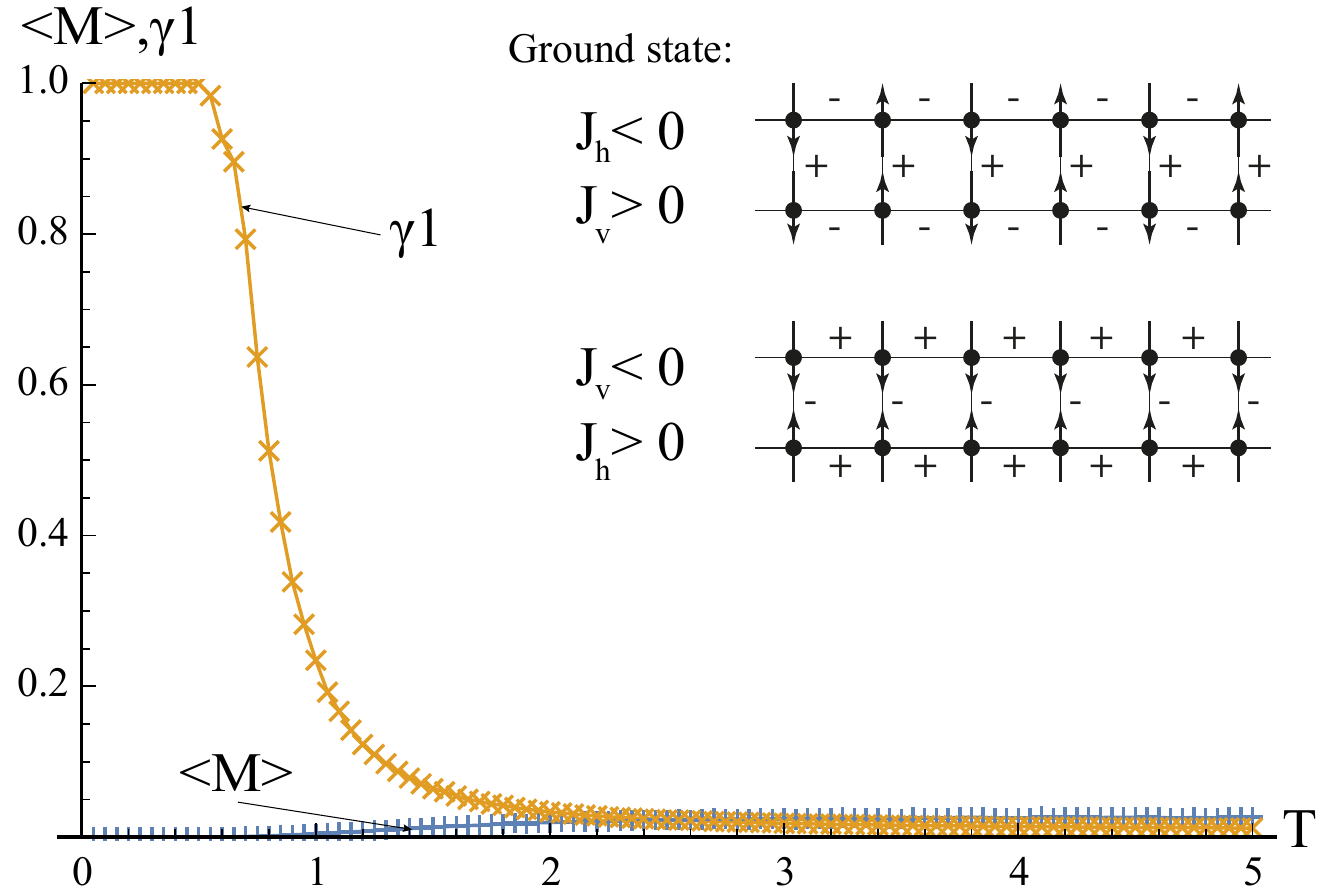}} \\ b)
\end{minipage}
\caption{The temperature behavior of COP $\gamma_1$ and magnetization of 1D Ising model (1000 spins) with three nearest neighbors and PBC. a) Blue denotes the magnetization of the ferromagnetic model with $J > 0$, green - magnetization of the antiferromagnetic model with $J < 0$, orange - COP (the same for both models). b) Blue denotes the magnetization with alternating exchange integral (each horizontal bonds of the spins are positive $J_h>0$, and the vertical are negative $J_v<0$ or vice versa), orange - the same COP $\gamma_1$ for such model.}
\label{Figure::1D3N_fig}
\end{figure}

Also, the 1D Ising model with $n=1000$ spins with three nearest neighbors and PBC (Fig. \ref{Figure::1D3N}) was investigated. As in the model with two neighbors, the ferromagnetic model (all exchange integrals are positive $J > 0$) shows the coincidence of the temperature behavior of the magnetization and COP $\gamma_1$ (Fig. \ref{Figure::1D3N_fig}a). At the same time in the model with $J<0$ the average magnetization is equal to 0, but COP $\gamma_1$ showed the transition to order.
As well as in model with alternating exchange integral (all vertical $J_v<0$, and horizontal $J_h>0$, or vice versa) COP $\gamma_1$ showed transition to order while magnetization is equal to 0 (Fig. \ref{Figure::1D3N_fig}b). 

\section{1D Ising model with four nearest neighbors}

\begin{figure}[h]
\begin{minipage}[h]{0.49\linewidth}
\center{\includegraphics[width=1\linewidth] {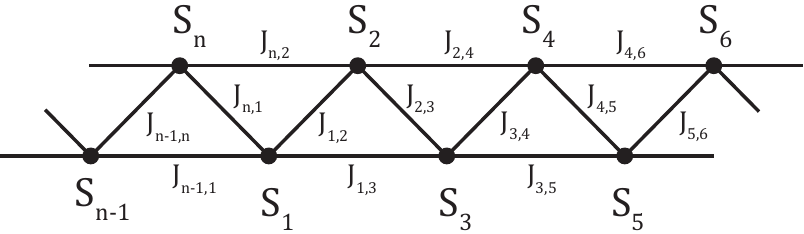}}
\end{minipage}
\hfill
\begin{minipage}[h]{0.49\linewidth}
\center{\includegraphics[width=0.6\linewidth]{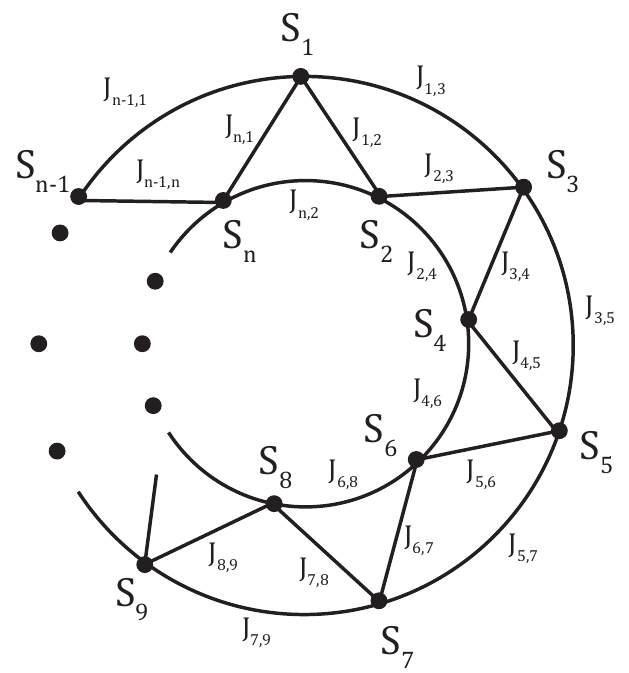}}
\end{minipage}
\caption{1D Ising model with four close neighbors and PBC.}
\label{Figure::1D4N}
\begin{minipage}[h]{0.49\linewidth}
\center{\includegraphics[width=1\linewidth] {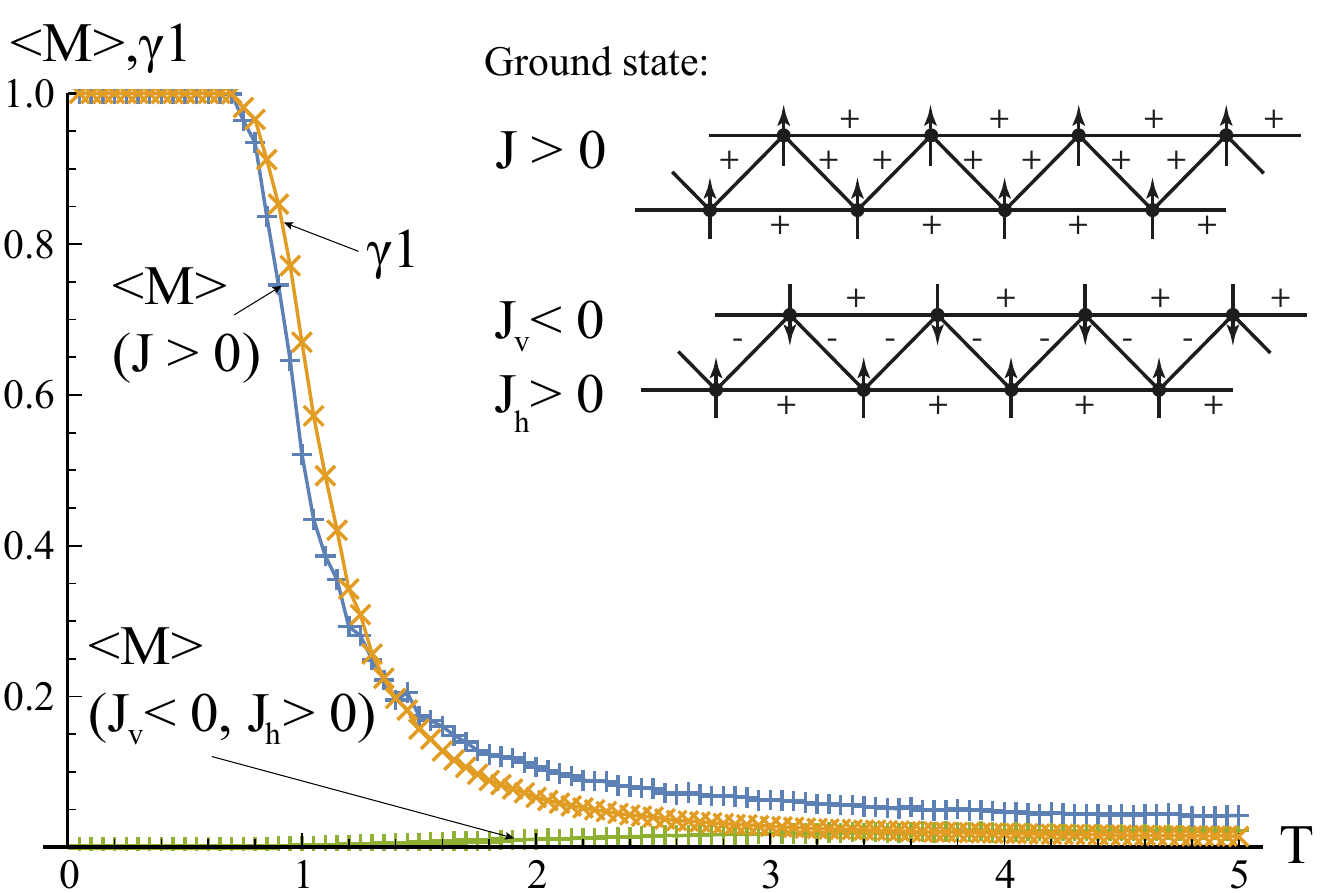}} \\ a)
\end{minipage}
\hfill
\begin{minipage}[h]{0.49\linewidth}
\center{ \includegraphics[width=1\linewidth] {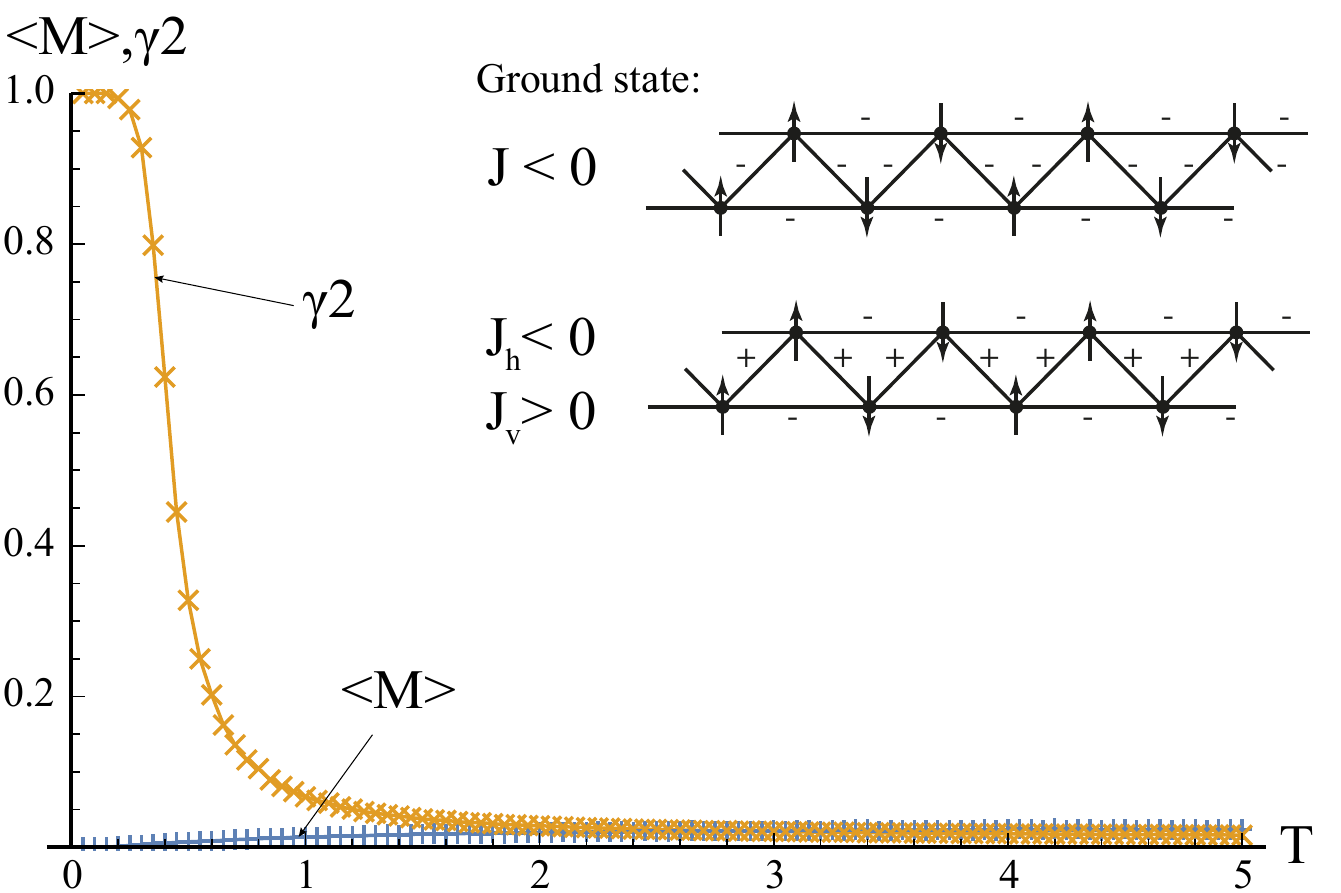}} \\ b)
\end{minipage}
\caption{The temperature dependencies of COP $\gamma_1,\gamma_2$ and magnetization of 1D Ising model (1000 spins) with four nearest neighbors and PBC. a) Blue denotes the magnetization of the model with $J > 0$, green - the magnetization of the model with $J_h>0$ and  $J_v<0$, orange - COP $\gamma_1$ same for these two models. b) Orange denotes COP $\gamma_2$ of frustrated systems with negative exchange integral $J<0$ and with alternating exchange integral with $J_h<0$ and $J_v>0$. Magnetization of these model is denoted as blue}
\label{Figure::1D4N_fig}
\end{figure}

The 1D Ising model with $n=1000$ spins, with four nearest neighbors and PBC (Fig.~\ref{Figure::1D4N}) was investigated. In case of ferromagnet interaction ($J > 0$) COP $\gamma_1$ coincides with the magnetization (Fig.~\ref{Figure::1D4N_fig}a). In model with alternating exchange integral $J_h>0$ and $J_v<0$ magnetization equal to 0, but COP $\gamma_1$ shows the ordering (the same for both model).

\begin{figure}[h]
\center{a) \\ \includegraphics[width=0.25\linewidth]{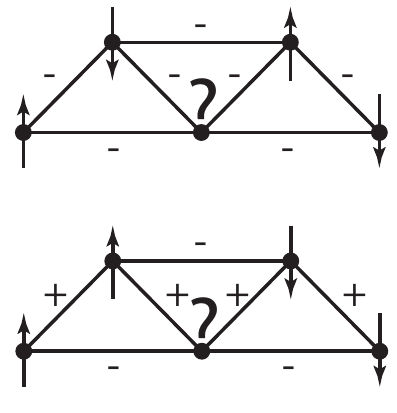} \\ b)    }
\caption{Frustrations in 1D Ising model with four nearest neighbors a) in the antiferromagnet model with $J < 0$ b) in the model with alternating exchange integral $J_h<0$ and $J_v>0$.  }
\label{fig:frustration1}
\end{figure}

But when we change the exchange integral to negative $J<0$ or to alternating with negative horizontal $J_h<0$ and positive vertical $J_v>0$ the frustration appears~(Fig.~\ref{fig:frustration1}).

The COP $\gamma_1$ does not working for frustrated systems because spins can not minimize its energy, which results to the impossibility of ground state cluster formation. However, under the proposed approach a more general clustered order parameter $\gamma_2$ for frustrated systems has been proposed.COP $\gamma_2$ represents a ratio of the maximum size of the cluster formed by spins not only in the ground state but also with energy closest to the ground state to the total number of particles (highlighted by the green line in Fig. \ref{Figure.cluster}).

Figure \ref{Figure::1D4N_fig}b shows that $\gamma_2$ describes ordering process in frustrating systems. The behavior of $\gamma_2$ is the same for two frustrated systems mentioned previously. Also, the clustered order parameter $\gamma_2$ can describe the ordering in other frustrated systems. 
A more detailed description of the approach can be found in the paper \cite{andriushchenko2013magnetic}.

\section{Conclusion}
In this paper, the investigation of one-dimensional and generalized one-dimensional Ising model was performed. All these data suggest that the COP is more universal than magnetization. In general, these models show similar results. Magnetization shows the transition only at positive $J$, but COP describes the transition, regardless the exchange integral sign and the number of neighbors.

The role of interacting spins can play the magnetic moments of nanoparticles as it is done in this work \cite{arnalds2016new}. Therefore, the experimental confirmation of the temperature dependences of the order parameter would be very interesting.

This work was supported the Ministry of Education and Science of the Russian Federation in the frame of the scholarship of the President of the Russian Federation for young scientists and postgraduate students performing advanced research and development in priority areas of modernization of the Russian Economics for 2015-2017 years (SP-1675.2015.5, order $\#184$, 10/03/2015).


\bibliographystyle{ieeetr}
\bibliography{citation1}

\end{document}